Published in "Ad hoc Networks" journal, 2020. DOI: https://doi.org/10.1016/j.adhoc.2020.102277# A Survey on Congestion Detection and Control in Connected Vehicles

Anirudh Paranjothi[1], Mohammad S. Khan[2], Sherali Zeadally[3]
[1]School of Computer Science, University of Oklahoma, Norman, OK, USA.
[2]Department of Computing, East Tennessee State University, Johnson City, TN, USA.
[3]College of Communication and Information, University of Kentucky, Lexington, KY, USA.*Abstract*—The dynamic nature of vehicular ad hoc network (VANET) induced by frequent topology changes and node mobility, imposes critical challenges for vehicular communications. Aggravated by the high volume of information dissemination among vehicles over limited bandwidth, the topological dynamics of VANET causes congestion in the communication channel, which is the primary cause of problems such as message drop, delay, and degraded quality of service. To mitigate these problems, congestion detection, and control techniques are needed to be incorporated in a vehicular network. Congestion control approaches can be either open-loop or closed loop based on pre-congestion or post congestion strategies. We present a general architecture of vehicular communication in urban and highway environment as well as a state-of-the-art survey of recent congestion detection and control techniques. We also identify the drawbacks of existing approaches and classify them according to different hierarchical schemes. Through an extensive literature review, we recommend solution approaches and future directions for handling congestion in vehicular communications.

**Keywords**— Congestion control, Congestion detection, Connected vehicles, VANET.## I. INTRODUCTION

Recent advancements in automotive communications have made connected vehicle technology a promising area of research in the field of transportation. Enabled by Dedicated Short Range Communication (DSRC), connected vehicles provide transformative solutions that ensure road safety and caters numerous transportation utilities enhancing the overall mobility experiences of travelers. These connected vehicles compose a special type of network, called Vehicular Ad hoc NETwork (VANET), which is a special form of Mobile Ad-hoc NETwork (MANET) with additional constraints [1, 2]. Connected vehicles communicate with each other using DSRC, providing support for various Intelligent Transportation System (ITS) applications and services such as 1) Road safety applications, 2) Infotainment services, 3) Messaging, and 4) Road-weather information. The primary goal of DSRC-based ITS applications is to ensure the safety of passengers by reducing the number of accidents. A plethora of safety applications have been developed using DSRC. Some well-known examples include forward collision warning, blind-spot warning, cooperative collision warning, intersection movement assistance, and lane change warning [3].

Connected vehicles also known as VANETs are equipped with two major communication devices: On-Board Units (OBUs) and Road Side Units (RSUs). OBUs are mounted inside vehicles while RSUs are placed at critical points of the road. Using OBUs, vehicles communicate with RSUs or other vehicles [4]. The Federal Communications Commission (FCC) had licensed a total of 75 MHz spectrum ranging between 5850 MHz and 5925 MHz for automotive communication [5, 6]. This spectrum is divided into seven channels, each spanning 10 MHz wide with a 5 MHz initial guard band. Among the seven channels, one is the control channel (channel 178), and the rest six are service channels as Figure 1 shows. The control channel transmits critical messages (i.e., roadblocks, road accidents, traffic information) to neighboring vehicles (i.e., vehicles in its transmission range). The service channels are used to transfer non-critical messages (i.e., entertainment information, personal messages, tolling information, and so on) to nearby vehicles [7-9]. DSRC uses Wireless Access in a Vehicular Environment (WAVE) for Vehicle-to-Vehicle (V2V) and Vehicle-to-Infrastructure (V2I) communications. WAVE is a set of standards defined by IEEE 802.11p for Physical (PHY) and Medium Access Control (MAC) protocols, and IEEE 1609.1 to 1609.4 for upper layer protocols [10].

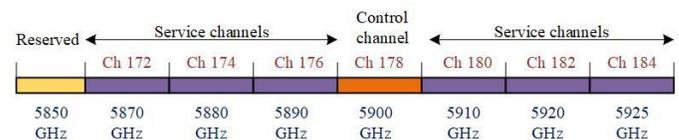

**Figure 1: DSRC communication channel.**

For VANET, a congestion control mechanism needs new approaches due to the unique characteristics of VANET such as high mobility, dynamic channel quality, and heterogenous devices [11]. To explain this through an example, let us take a generic scenario for connected vehicles in urban and highway environments as Figures 2 and 3 show respectively, where vehicles are connected to RSUs. When the number of vehicles connected to an RSU increases, it leads to increased channel contention at the RSU, which in turn leads to problems such as increased latency and possible packet loss which degrade the overall performance of the network. Congestion mechanism characteristics are based on different criteria such as congestion detection and control, hop-by-hop and end-to-end, passive and active, Media Access Control (MAC), network and transport layer and cross-layer approaches. In a congestion detection mechanism, when a packet loss occurs in the network, then congestion is detected in the network. Then, the network takes measures to recover from the congestion. On the other hand, a congestion control mechanism attempts to take measures to handle the congestion before its occurrence by using various strategies. Moreover, an effective and efficient mechanism for congestion detection and control also enhances the Quality of Service (QoS) in VANET, which is primarily governed by two factors: 1) packet loss and 2) delay [12-14]. Existing congestion detection and control strategies employed in VANETs focus on three main objectives: 1) Controlling the transmission range, 2) Controlling the transmission rate, and 3) Priority scheduling [15-17]. Transmission range is controlled by varying the transmission power. The transmission rate is used to control the rate of packet transmission. Priority scheduling is used to schedule messages in multiple channels based on priorities [18].

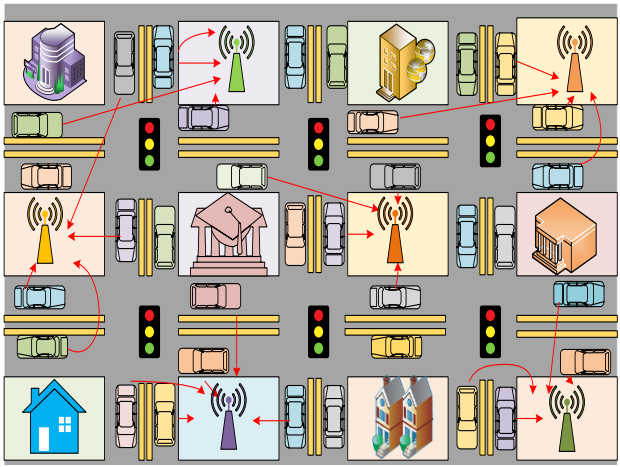

**Figure 2: Generic VANET scenario – urban environment.**

In this work, we classify these strategies into two categories: 1) Congestion detection, 2) Congestion control. We discuss existing congestion detection and control techniques used in VANET, parameters used in existing approaches, drawbacks of existing approaches and we propose a solution for congestion control in VANET along with some recommendations on future directions. Figure 4 represents a scope and detailed taxonomy of our survey paper.

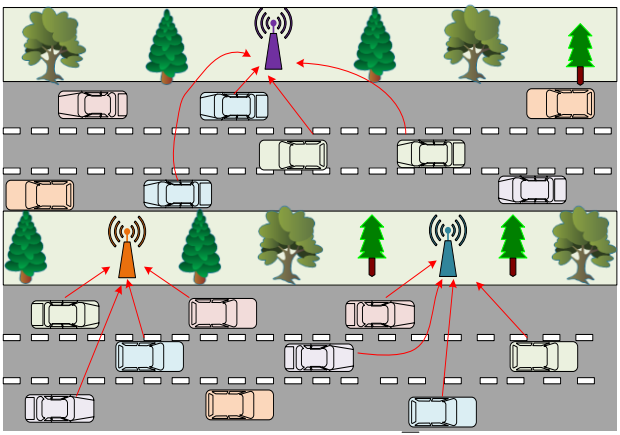

**Figure 3: Generic VANET scenario – highway environment.**

The contributions of this paper are as follows:
- We have explored existing congestion detection and control techniques in connected vehicles, and classified the proposed algorithms into three main categories: 1) reactive algorithms, 2) proactive algorithms, and 3) hybrid algorithms.
- We present an extensive review of congestion detection and control techniques, including the advantages, limitations, and complexities of the proposed algorithms.
- We analyze some open issues and recommend possible solutions to overcome the limitations of the existing congestion detection and control techniques.

The rest of the paper is organized as follows: Section II discusses the related surveys in both congestion detection and control strategies and also compares DSRC-based congestion control with other communication technologies. Section III and IV describe congestion detection and control strategies respectively. Section V focuses on the challenges of existing congestion detection and control schemes and outlines future research directions. Finally, section V makes some concluding remarks.

## II. RELATED WORKS

In VANETs, vehicles communicate with each other by transmitting and receiving messages. If a vehicle encounters a situation like an accident or traffic congestion, event-driven messages are generated and transmitted to all the vehicles in the region. All these messages are time-critical and should reach the destination within a specific time interval. Over the last decade, researchers have proposed various congestion detection and control techniques to monitor, detect, and mitigate congestion in the network to enable better use of bandwidth and provide higher QoS. VANETs pose a problem for congestion control, due to the need to match a variable workload to an inherently unstable network topology. Strategies for VANET congestion detection and control can be divided into those that address congestion after it occurs and those that address congestion before it occurs. This section discusses the recent congestion control surveys available in the literature and compares DSRC-based congestion control with other communication technologies.

### 2.1 Comparison of existing surveys on DSRC-based congestion control

Liu et al. [19] presented a comprehensive survey illustrates the importance of congestion control in VANETs and classifies existing congestion control techniques based on performance metrics such as channel capacity, delay, and bandwidth utilization. However, the paper discussed only decentralized V2V congestion control techniques in VANETs. Also, there are no discussions about drawbacks of the various congestion control techniques given in the literature. Taleb et al. [20] proposed a congestion control survey based on location and MAC-based techniques. The paper discusses various parameters that need to be considered to reduce congestion in the network. The authors [20] discussed only three congestion control algorithms in the survey paper. Moreover, the taxonomy, challenges, and future directions of the existing congestion control techniques were not discussed in the paper.

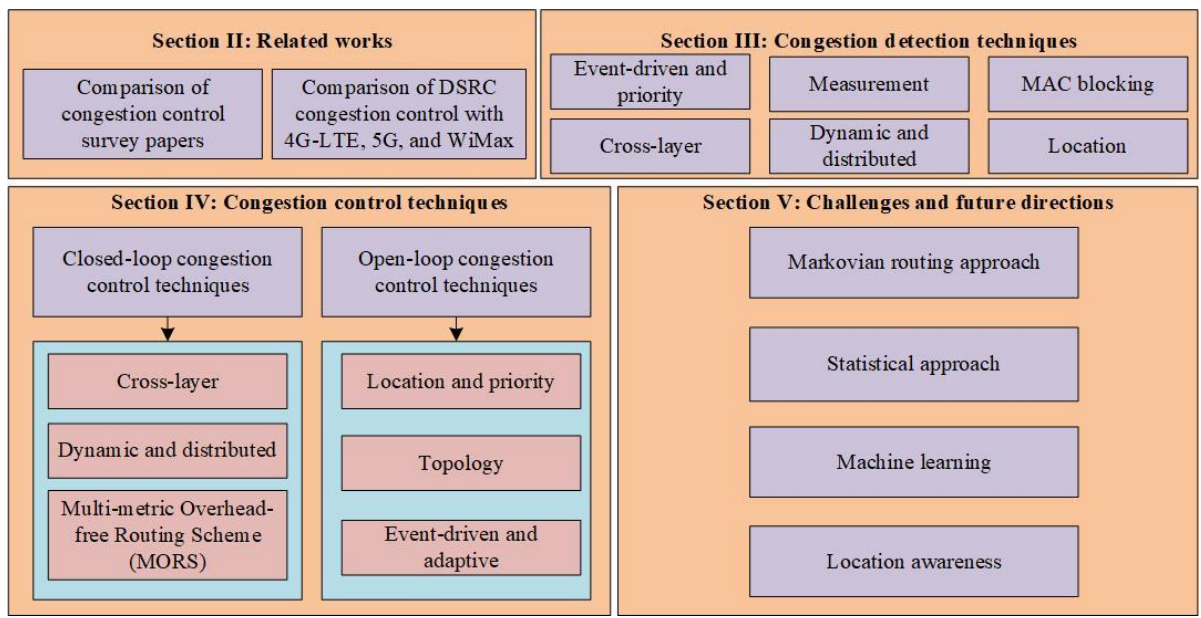

**Figure 4: Scope and taxonomy of this survey paper.**

Elias et al. [21] have presented an overview of VANETs congestion control algorithms that minimizes congestion by altering the transmission power and packet generation rate. As in [20], the authors of [21] also discussed only three congestion control algorithms in the survey paper. Moreover, they did not provide any insight into the challenges and future directions of the congestion control techniques. Song et al. [22] surveyed decentralized congestion control techniques for VANETs. The authors discussed the congestion control techniques based on the following assumptions: 1) All vehicles use the DSRC technique for communication, 2) Control channels are shared by all vehicles in the region, and 3) Critical messages in VANETs have higher priorities compared to non-critical messages. The paper also presented a classification of decentralized congestion control based on the IEEE 802.11p MAC protocol. However, the advantages, limitations, and challenges of various decentralized congestion control techniques were not discussed in the paper.

Jarupan et al. [23] reviewed cross-layer congestion control techniques for connected vehicles. The paper presents an overview of cross-layer congestion control based on physical, MAC, network, transport, and application layer using the following performance metrics: 1) Implementation strategy, 2) Message transmission and reception rate, 3) Channel selection and 4) Channel usage. However, they did not discuss the various congestion control techniques used considering characteristics such as dynamic, distributed, and location-based solutions. Also, the did not discuss future directions of cross-layer congestion control. Nahar et al. [24] surveyed MAC layer-based congestion control techniques in VANETs and presented various challenges involved in various congestion detection and control techniques. The authors also illustrated the effectiveness of various parameters, such as message transmission rate and power rate in reducing congestion across the region. However, the survey focused only on decentralized congestion control techniques in the MAC layer. But the authors did not present a detailed taxonomy and they did not discuss the limitations and future directions of the congestion control techniques they have reviewed in that work.

To address the limitations of existing survey papers available in the literature and to provide an extensive overview of congestion detection and control techniques, in this survey paper, we have categorized and discussed existing congestion detection and control in VANETs based on event-driven, priority, measurement, MAC, cross-layer, distributed, location, open-loop, and closed-loop based techniques. We have evaluated various congestion control techniques based on delay, overhead, packet loss, energy, and mobility. Moreover, our survey paper also presents a detailed taxonomy, challenges, and future directions to help the readers better understand the pros and cons of various congestion control techniques and their limitations.

**2.2 Comparison of DSRC-based congestion control with other communication technologies**

This subsection compares and analyzes DSRC-based congestion control in VANETs with various communication technologies that are used. These wireless technologies include cellular networks (4G-Long Term Evolution (LTE)), 5G, and Worldwide Interoperability for Microwave Access (WiMAX). We have also summarized the list of auto manufacturers who are still using DSRC technology for V2V and V2I communications to demonstrate the relevance of DSRC today.

*i) Cellular Networks (4G-LTE):* Cellular networks, in particular, 4G-LTE has the potential to revolutionize the VANETs due to its characteristics such as low latency, high bandwidth, and high throughput. 4G-LTE cellular networks operate on the frequency of 1.9 GHz, comprises the following components to establish a communication between the connected vehicles: 1) Customer Premise Equipment (CPE) terminals, 2) eNode-B base station, 3) Evolved Packet Core (EPC), 4) Server clusters, and 5) Switches [25]. OBU connects to LTE networks through CPE, which transmits the messages to destination vehicles through wired or wireless networks. The eNode-B is responsible for the allocation of resources, packet scheduling, bandwidth, and mobility management. EPC consists of various gateways responsible for data processing and exchange. 4G-LTE cellular networks work with four cells, and each cell can provide a maximum uplink rate of 20 Mb/s and a downlink rate of 80 Mb/s, even at high vehicle densities [26]. Moreover, 4G-LTE networks work well for both urban and highway scenarios due to its high bandwidth and throughput. Therefore, of the impact of network congestion is lower compared to DSRC technology. DSRC technology heavily adopts RSUs for communication. RSUs are resource-constrained and thus, there is a high chance of network congestion compared to the 4G-LTE platform resulting in need of efficient congestion control techniques to minimize the congestion in the network.

*ii) 5G-networks:* 5G vehicular communication is an emerging platform and gained the attention of both academia and industry due to the plethora of novel applications responsible for providing ultra-low latency communication [27]. 5G networks do not change the current LTE network architecture, instead, they provide a platform, which can leverage various existing techniques of the 4G-LTE platform. As a result, 5G provides very high bandwidth and greater coverage area for the device to device (D2D) communication [28, 29]. To provide ultra-low latency and efficient bandwidth utilization for connected vehicles, 5G networks exploit the mobile edge computing (MEC) technique, which provides services at the proximity of the users. Moreover, User Equipment (UE) in 5G yields high data transmission rates and efficient resource utilization. As a result, 5G significantly reduces the chances of congestion in the network. Therefore, congestion detection and control in 5G is not a major research issue.

In contrast, the major portion of the DSRC spectrum allocated for periodic beaconing results in high utilization of channel load at high vehicle densities. RSUs depend on DSRC communications in VANETs to transmit and receive messages and they are deployed only in the critical regions of VANETs. Thus, there is a high chance of congestion at vehicle dense regions. To cope up with the congestion, the DSRC must incorporate congestion control techniques based on either open-loop or closed-loop strategy. The congestion control techniques perform load balancing, modify transmission rate, and so on to reduce the congestion in the network.

*iii) WiMAX:* WiMAX is a wireless communication technology designed to provide low latency and high data rate for connected vehicles. It consists of a WiMAX tower and receiver antenna responsible for vehicular communication. WiMAX consists of two standards: 1) IEEE 801.16d and 2) IEEE 802.16e. IEEE 801.16d used for fixed stations with a data rate of 70 Mb/s and a coverage range of 48 Km. IEEE 801.16e used for mobile nodes, including connected vehicles with a data rate of 10 Mb/s and a coverage range of 10 Km [30]. Although WiMAX has some benefit such as low delay for small packets over DSRC, the limitations of WiMAX include: 1) Expensive network with high installation and operation cost, 2) Poor QoS at heavy traffic conditions, 3) Poor bandwidth allocation with 2 to 10 Mb/s of shared bandwidth, 4) High power consumption, and 5) Interference problems. Moreover, the emergence of novel paradigms such as Mobile Edge Computing (MEC), fog computing, cellular (4G-LTE), and 5G networks has decreased interests of researchers and industry in the WiMAX technology. Therefore, no novel congestion detection and control techniques have been published recently in the literature.

According to Forbes magazine, Volkswagen introduced DSRC-based vehicle Vehicle-to-Everything (V2X) technology with the new launch of the eighth-generation Golf using a new chipset from NXP. The chipset enables V2V, V2I, and V2X connectivity. The DSRC-enabled Golf can send emergency information (such as slippery road, brake failures) to nearby vehicles in an 800 m radius [31]. Additionally, emergency vehicles in Europe are already equipped with DSRC technology to broadcast emergency messages among vehicles. The report from the GM and Cadillac illustrates a 5.9 GHz spectrum allocated by FCC for DSRC-based vehicular communication [32, 33]. Cadillac DSRC technique can handle 1000 messages per second from vehicles up to 1000 m radius. GM was the first auto industry in the U.S. market to deploy DSRC-based V2V connectivity for its Cadillac CTS in 2017. DSRC is SIM card-free Wi-Fi whereas 5G-V2X cannot exist without a SIM card. The operational costs of 5G-V2X remain unknown.

The performance of 4G-LTE networks is worse compared to DSRC technology in the collision avoidance scenario caused by the Doppler Effect. Thus, the 4G-LTE platform is not suitable for safety-related applications [25]. Moreover, the United States Department of Transportation (US-DOT) is still using DSRC-based V2X communication, and there are 70 deployments with thousands of vehicles that are already on the road [34]. The auto industries and government agencies are still utilizing DSRC technology to enable vehicular communications. Therefore, to provide better QoS, DSRC should be equipped with efficient congestion control techniques based on either open-loop or closed-loop congestion control strategy. Similar to Figure 1 in [35], in our paper, Figure 5 represents the congestion scenario of the DSRC-based technique.

### III. CONGESTION DETECTION TECHNIQUES

Congestion detection mechanisms in VANET allow collecting information of congested links at a time interval (*t*). Once congestion is detected, control techniques are applied to mitigate the congestion. The congestion detection section in this paper classifies existing congestion detection techniques in connected vehicles based on six major classification such as measurement-based congestion detection technique, event-driven and priority-based congestion detection technique, Media Access Control (MAC) blocking-based congestion detection technique, cross-layer-based congestion detection technique, dynamic and distributed-based congestion detection technique, and location-based congestion detection technique. The measurement-based congestion detection technique senses communication channels and measures parameters such as the number of messages in the queue, channel usage level, and channel occupancy time [36-40]. The congestion detection component measures the channel usage level periodically to detect any congestion situation.

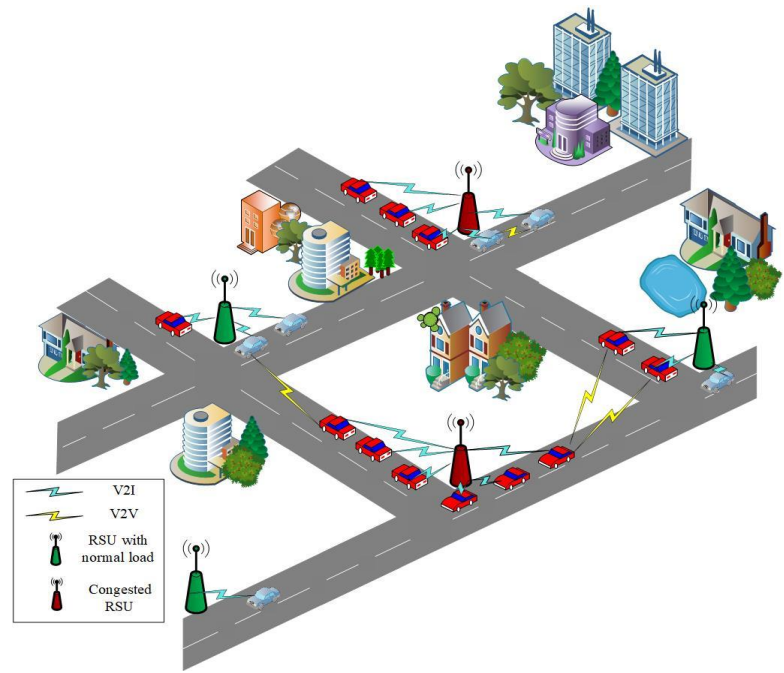

**Figure 5: A congestion scenario of the DSRC-based technique.**

Event-driven detection technique monitors the event-driven safety message and decides to start the congestion control algorithm whenever an event-driven safety message is detected or generated [41, 42]. The MAC blocking detection technique is based on the control of beacon message transmissions to reduce the congestion and traffic rate control for congestion avoidance. The cross-layer approach detects congestion at various layers of the network. Dynamic and distributed congestion detection includes parameters such as channel usage level, channel occupancy time and the number of messages in a queue. In location and priority-based approaches, the congestion path of intersections is determined based on the normalized length of the path and the connectivity of the path. For congestion control, Dijkstra's shortest path algorithm is used to determine the optimal path [43]. Table 1 presents a summary of congestion detection techniques.

### 3.1. Event-driven and priority-based congestion detection technique

Event-driven and priority message congestion detection techniques start a congestion detection algorithm whenever event-driven messages about traffic jams, road accidents are detected or generated. Fadilah et al. and Biswas et al. [44, 45] proposed the use of adaptive periodic beaconing to convey time gap data. The adaptive rate control algorithm varies the rate of Periodic Safety Message (PSM) generation based on vehicular safety. PSMs can cause packet loss and increase the busy channel percentage due to the short Control CHannel (CCH) interval. This creates safety risks for vehicles. The proposed congestion detection scheme identifies the congested node when a packet loss is greater than a threshold value. A distance-based wait time method selects rebroadcast nodes and retransmits messages to mitigate multipath fading in multi-hop networks. Bouassida et al. [46] proposed a congestion control approach based on the properties of dynamic priority-based scheduling. The congestion control algorithm detects congestion based on network load and priority messages transmitted through control and service channels. As the priority scheduling reduces the congestion of the network, the end-to-end delay of high priority messages is significantly lower compared to low priority messages. However, these approaches [44-46] suffer from high routing overhead in areas where the densities of vehicles are high.

In [47, 48], Taherkhani et al. and Feukeu et al. proposed a dynamic and distributed strategy for congestion detection for VANETs. This strategy also detects congestion. Congestion is detected by sensing the channel usage level and comparing it with a predefined threshold, set to 70% in wireless communication channels. Thus, the channel never gets congested even when the number of vehicles increases in the region. However, due to the predefined threshold, the packet loss increases at high vehicle densities and in downtown areas. In [49], Pierre proposed two congestion detection strategies to prioritize and schedule safety and service messages. These strategies use a priority assignment unit, which prioritizes messages based on static and dynamic factors and message size, together with a message scheduling unit, which reschedules control and service channel queues before transferring the messages to channels. These priorities are embedded in packet headers. The end-to-end delay of the congestion strategy described in [49] is higher because of the overheads caused by rescheduling control and service channel queues.

### 3.2. Measurement-based congestion detection technique

The measurement-based congestion detection technique detects congestion based on channel capacity and channel usage. The channel capacity and channel usage are compared with the predefined or target value to determine if the channels are congested [50, 51]. Zang et al. [52] proposed a technique for congestion detection which monitors and compares the channel usage with a predefined threshold value at a constant time interval (*t*). When the channel usage exceeds a predefined threshold limit, congestion is detected, and the information of congested channels is transmitted to all the vehicles in that region. He et al. [53] illustrated the DSRC-based congestion detection technique for vehicular networks. The authors concluded that when the channels are heavily congested, more than 70% of the messages are dropped in the middle without reaching the destination.

Taherkhani et al. [54] proposed a congestion detection based on meta-heuristic techniques. In this approach, two units such as the detection unit and measurement unit are designed for congestion detection. The measurement unit checks the channel usage, number of messages in a queue, and channel occupancy time. A channel usage of more than 70% is considered as the likelihood of congestion and notifies the detection unit. The detection unit closely monitors the channel and broadcasts congestion information to all the vehicles when the channel usage exceeds the target value. The benefits of the approaches [52, 54] are low overhead and low packet loss. But the end-to-end delay of the dynamic congestion control significantly increases when the number of vehicles increases because a large number of messages need to be monitored and analyzed to detect congestion in the network.

### 3.3. MAC blocking-based congestion detection technique

The MAC blocking-based congestion detection technique detects congestion on links and channels based on beacon message transmission rate and traffic rate [55]. We discuss the existing MAC blocking-based congestion detection techniques in this subsection. Bellache et al. [56] proposed a proactive Contention-Based Forwarding (CBF2C) mechanism strategy to detect congestion while providing transmission redundancy whenever possible. The strategy uses the busy channel ratio to determine a retransmission threshold and traffic rate and does timed retransmissions until a retry threshold is reached. Moreover, the CBF2C algorithm monitors the channel load status. Based on this load it gathers the information and determines congested links and congestion channel in that region. Willis et al. and Rath et al. [57, 58] proposed a congestion detection scheme that monitors the transmissions to nearer vehicles over transmissions to more distant vehicles. Their scheme balances adjustments to transmission power and transmission rate to minimize the congestion.

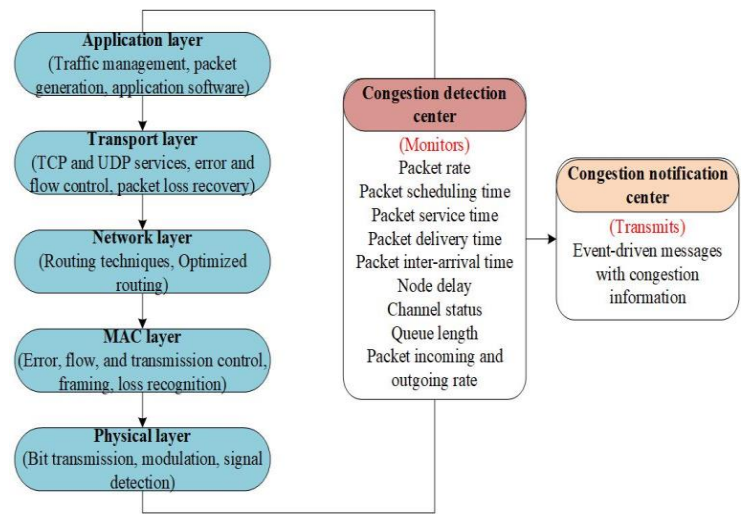

Figure 6: Cross layer congestion detection.

Math et al. [59] focus more on packet count than specific threshold values. Their Packet-count based Decentralized data-rate Congestion Control algorithm (PDR-DCC) algorithm detects channel congestion by monitoring the maximum permissible data rate and traffic rate that ensures maximum coverage. The congestion is detected when the data rate and traffic rate exceed the threshold value. Chen et al. [60] proposed a congestion detection scheme based on a non-cooperating bargaining game. The game divides vehicles into clusters, each of which has a leader. The clusters' leaders negotiate with each vehicle for an optimal combination of transmission power and packet generation rate for their members. Ideally, negotiations should yield an equilibrium state that maximizes each player's utility to reduce congestion. This approach experiences high end-to-end delay caused by the initialization and formation of clusters.

### 3.4. Cross-layer-based congestion detection technique

This approach focuses on critical messages to ensure the safety of passengers and drivers. These safety messages should be carried to the neighboring vehicles without any delay. To accomplish this, a high bandwidth of communication channel is utilized [61].

A cross-layer congestion detection scheme consists of two modules. The first module is responsible for event-driven message detection. This module scans for an emergency and alerts a control center if any emergency message gets delivered. The second module channel senses the overall load of the channel [62]. Sensing is based on the assessment of dynamic threshold values, queue length, packet rate, scheduling time, delivery time, incoming and outgoing rate to determine congestion. Using this technique congestion at a particular channel can be identified. To utilize the bandwidth channel in an appropriate way dynamic threshold values are used instead

**Table 1: Summary of congestion detection mechanism based on different parameters**

| Approach | Message type | Congestion detection methods | Limitations |
|---|---|---|---|
| Cross-layer approach [61] | Safety message | Event-driven messages | High delay and jitter |
| Dynamic approach [47] | Beacon message | Measurement-based | High communication overhead |
| Distributed approach [63, 64] | Beacon message | Event-driven messages | Channel congestion |
| Cooperative vehiculAR Traffic congestion Identification and Minimization (CARTIM) [65] | Beacon message | Measurement-based | High packet loss |
| Decentralized approach [66, 67] | Safety message | MAC blocking detection | High delay and packet loss |
| Data mining approach [68] | Beacon message | Event-driven messages | Channel congestion |
| Dynamic distributed approach [69] | Beacon message | MAC blocking detection | High delay |
| Cross-layer coordination of multiple vehicular protocols (COMPASS) [70] | Beacon message | Event-driven messages | High communication overhead |
| Periodically Update Load Sensitive Adaptive Rate control (PULSAR) [71] | Safety message | Measurement-based, Event-driven messages | High packet loss and jitter |
| DSRC based congestion control [72] | Beacon message | MAC blocking detection | High delay |
| Adaptive Beacon Generation Rate (ABGR) congestion control [41] | Beacon message | Event-driven messages | High delay and packet loss |
| Location based approach [73] | Safety message | Measurement-based | High packet loss and jitter |

of a predefined one [36]. Figure 6 shows the cross-layer congestion detection model.

### 3.5. Dynamic and distributed approach-based congestion detection technique

A dynamic and distributed approach is also known as MOTabu for congestion detection [47]. Congestion detection is based on the channel usage. The performance of the MOTabu approach is measured based on the urban and highway scenario and it considers five parameters: 1) average delay, 2) number of packets lost, 3) average throughput, 4) packet loss ratio, 5) number of retransmissions.

Congestion is detected by using predetermined methods. These methods scan channels and measures parameters such as channel usage level, channel occupancy time and the number of messages in a queue [63]. This approach has considered only channel usage level as a parameter for congestion detection. The channel usage level is calculated periodically and compared with a predefined threshold value [64]. The predefined threshold value is 70% usage of a wireless communication channel. If the channel usage exceeds this capacity (i.e.,70%), then the channel will be considered congested, and congestion control mechanism will be applied to that channel to reduce congestion. Once congestion is detected, congestion control techniques are applied to minimize the congestion. A tabular search algorithm is used in congestion control. This algorithm is used to mitigate the congestion occurring on a particular channel.

### 3.6. Location-based congestion detection technique

Routing of messages in the connected vehicular environment is challenging with high vehicle densities, which leads to frequent disconnection problems due to the service channels responsible for the transmission of messages that are heavily congested [74, 75]. When the number of vehicles increases in a region, channel usage increases as a larger number of messages need to transmit among the vehicle in a specific time interval (*t*). For example, in a high dense vehicular environment, such as Manhattan and other downtown regions frequently encounter congestion due to the unavailability of service channels for transmitting messages within a region. The existing location and priority-based approach for congestion detection determines the congestion link, congestion node, and congestion channel based on normalized length and connectivity of the path.

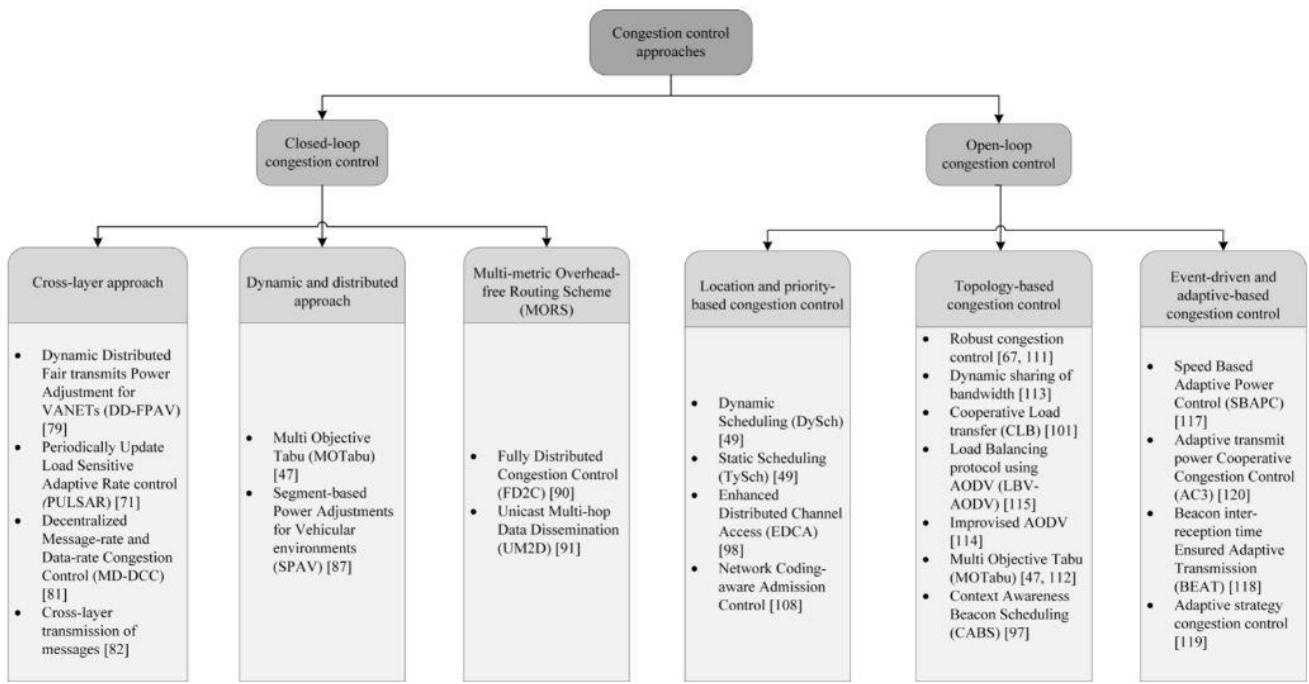

**Figure 7: Congestion control approaches based on detection methods.**

In the location-based routing technique [76, 77], the congested route with a sequence of intersections is determined based on statistical techniques. When a link or channel is congested, the RSU sends the information of the congested link or channel to the location server, which returns the congested route with a sequence of intersections. This information is then transmitted to the set of vehicles located in that region and the information is saved in the routing table. The location server determines the congested link or channel based on the length of the path and connectivity of the path. The shortcomings of these approaches are high delay, routing, and communication overheads. These approaches require high cooperation and communication among vehicles and RSUs to detect and deliver the congestion information to other vehicles located in that region.

In Section III, we have discussed state-of-the-art congestion detection techniques based on event-driven, priority, measurement, MAC blocking, cross-layer, dynamic, distributed, and location-based techniques, out of which, the cross-layer and location-based congestion detection techniques gained the most attention among researchers in recent times. The cross-layer congestion detection mechanism [61, 62] monitors all TCP/IP layers to detect network congestion. In contrast, location-based congestion detection techniques provide statistical techniques to detect congestion in the region. The centralized controller, such as RSUs provide efficient path to all vehicles. Moreover, cross-layer and location-based congestion detection techniques work efficiently for regions with high vehicle densities. Thus, the cross-layer and location-based congestion detection algorithms are heavily used compared to all other congestion detection techniques. Some of the cross-layer and location-based congestion control techniques require high cooperation among vehicles and therefore incur high overheads and delays in high vehicle density areas. To overcome these limitations, we have recommended some important techniques to perform congestion detection in Section V.

## IV. CONGESTION CONTROL TECHNIQUES

When the number of vehicles communicating at a given time (*t*) in a VANET increases, the challenges for congestion control also increase. To mitigate the congestion that occurred at a particular channel and maintain the channel under a predefined threshold value, various congestion control schemes have been proposed. This section briefly discusses these existing congestion control mechanisms used in VANET. The congestion control schemes can be classified using their features, such as topology, adaptiveness, the simulator used, the channel used, reactiveness, and performance metrics. Congestion can be controlled before it occurs, or a control mechanism can be implemented based on the occurrence of congestion in the network.

Congestion control mechanisms are broadly classified into two categories: 1) Closed-loop congestion control and 2) Open-loop congestion control. An open-loop congestion control mechanism prevents congestion from occurring and a closed-loop congestion control scheme controls the congestion after it occurs based on pre-congestion or post-congestion strategies. Figure 7 depicts a detailed classification of various algorithms based on their open-loop or closed-loop characteristics.

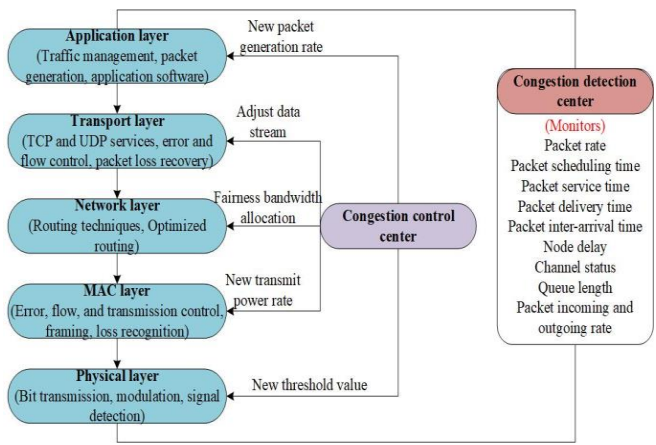

**Figure 8: Congestion Control in cross-layer approach.**

### 4.1 Closed-loop congestion control

In closed-loop congestion, congestion control mechanisms are proposed after the congestion is detected. We discussed three main approaches used in closed-loop congestion control environment. 1) Cross-layer approach, 2) Dynamic and distributed approach and, 3) Multi-metric Overhead-free Routing Scheme (MORS). The congestion detection mechanism of these approaches is discussed in the previous section. Table 2 presents a summary of closed loop congestion strategies.

#### 4.1.1 Cross-layer approach

A cross-layer approach focuses on dynamic load balancing, which results in de-normalizing congestion that occurred at a channel. The congestion control takes place at all networking layers as follows [61]. The application layer uses various methods (i.e., condition-based and application-based) for congestion control, and controls the generation of packets. At the transport layer, User Datagram Protocol (UDP) is used to broadcast the packets. The network layer uses various algorithms such as artificial intelligence algorithms, routing algorithms, broadcasting algorithms [78] to reduce the channel load and congestion. At the MAC layer, providing priority to the packets is the basis of congestion control. Packets with lower priority are dropped to reduce the channel load. The physical layer provides the first step of congestion control. At this layer, congestion on the channels is detected by monitoring and assigning predefined values to it. Dynamic Distributed Fair transmits Power Adjustment for VANETs (DD-FPAV), PULSAR, Decentralized Message-rate, Data-rate Congestion Control (MD-DCC), and Cross-layer-based transmission of messages are the most commonly used algorithms in the cross-layer approach. Figure 8 shows the cross-layer congestion control approach.

*i) DD-FPAV:* The DD-FPAV algorithm's congestion control goal is achieved by controlling the packet generation rate and transmission power [79]. In this algorithm, the channel load (i.e., threshold value) is calculated considering diverse road conditions such as high traffic and low traffic conditions and event-driven messages. High and low traffic conditions are distinguished based on the information carried by beacon messages, and event-driven messages are identified based on a special flag used in received messages. After identifying channel load, Maximum Beaconing Load (MBL) and Beacon Generation Rate (BGR) are selected. MBL is calculated based on the dynamic clustering algorithm [79]. The advantages of the DD-FPAV algorithm are low bandwidth utilization and low overhead. However, this approach encounters packet loss when the channel load exceeds the predefined threshold value.

*ii) PULSAR*: PULSAR is a reactive approach, where congestion control mechanisms are applied after the congestion is detected [71]. In PULSAR, the vehicles measure Channel Busy Ratio (CBR) at every time interval (*t*). When the measured CBR is higher than the target value, congestion is detected. Congestion control approaches are applied to reduce congestion. The time interval between CBR measurements is known as Channel Monitoring and Decision Interval (CMDI), a fixed interval for all vehicles [80]. The congestion control mechanism is responsible for reducing the transmission range to the target value. To eliminate the congestion, the authors applied a congestion control mechanism in PULSAR, which reduces the transmission range based on the required target range and adapts the transmission range of CBR. The limitations of PULSAR include high delay and high energy consumption until the channel loads reduce to the target value.

*iii) MD-DCC*: MD-DCC is one of the effective congestion control schemes to reduce congestion among connected vehicles. The algorithm provides an optimized and efficient way of message rate and data rate among the vehicles to minimize congestion on channels [81]. In MD-DCC, the authors declared the frequency of beacon messages to be lower than the required value by reducing the message rate to minimize congestion. It is suitable only for less dense vehicle regions. When the density of vehicles increases (for example, in an area such as the downtown environment), the number of vehicles is higher compared to the urban environment. In such cases, the MD-DCC algorithm dynamically adjusts the data rate for more channel bandwidth to avoid congestion. The shortcomings of the MD-DCC approach are: 1) As the vehicles transmit messages at a different data rates, synchronization could be a major problem between the sender and the receiver, 2) High vehicle density regions require high data rate resulting in a high Signal to Noise Ratio (SNR).

*iv) Cross-layer-based transmission of messages*: In this approach [82], the authors illustrated Enhanced AODV (EAODV) algorithm to minimize the congestion in VANETs. The EAODV algorithm monitors the message reception rate, channel load, and bandwidth utilization ratio to detect congestion in the network. When congestion occurs, the EAODV algorithm calculates the optimum message transmission ratio and transfers a load the from the service channel queue to the control channel queue to minimize the congestion. The EAODV algorithm is used for both low and high traffic conditions. However, the limitations of cross-layer-based transmission of messages include high overheads and high delays arising from the exchange of messages between the service channel queue and the control channel queue.

**Table 2: Summary of decentralized closed-loop congestion control methods based on different parameters using both pre- and post-congestion techniques.**

| Approach | Adaptiveness | Simulator used | Channel used | Performance metrics |
|---|---|---|---|---|
| Meta-heuristics [54] | - | Adaptive Network Simulator (NS2) | Channel switching and enhance reliability | Event and Measurement driven |
| Uni-Objective Tabu (UOTabu) and MOTabu [47] | - | SUMO, MOVE, and NS2 | CCH and SCH | Event-driven and measurement driven |
| DD-FPAV [79] | - | SUMO, MOVE, and NS2 | CCH | Event-driven |
| Cross-layer approach [61] | Adaptive | WARP2 | CCH and SCH | Event and measurement driven |
| Rebroadcast algorithm [69] | - | OMNET++, SUMO | CCH and SCH | Event and measurement driven |
| Tabu search [47] | - | SUMO | CCH and SCH | Event-driven |
| Dynamic and distributed approach [86] | - | - | CCH and SCH | Event-driven |
| PULSAR [71] | Adaptive | - | CCH and SCH | Event-driven |
| enhanced Multimedia Broadcast Multicast Services (eMBMS) technique [64] | Adaptive | VISSIM and NS3 | CCH | Event-driven |
| Multistate-active DCC [66] | Adaptive | PHY-MAC | CCH | Event-driven |

*4.1.2 Dynamic and Distributed Approach*

In a distributed approach, the two main factors. packet loss, and delay are considered for congestion control. A congestion control strategy for a distributed approach works as follows: it minimizes delay and jitter which in turn provide the flexibility to control the transmission range and the transmission rate. It is a challenging task due to the frequent topology changes and mobility of nodes. This mechanism helps in controlling congestion among connected vehicles. But there are some drawbacks in using this mechanism. It increases the message collision rate when the communication range and the transmission rate increase. Hence, the optimal value needs to be used in both the transmission rate and the transmission range. MOTabu and Segment-based Power Adjustments for Vehicular environments (SPAV) are commonly used distributed congestion control algorithms.

*i) MOTabu:* The MOTabu algorithm is comprised of components such as an initial solution, objective functions, searching strategy, memory mechanisms, terminating conditions, and tabu list [47]. The performance of this algorithm is based on length of the tabu list. The elements of the MOTabu algorithm should be mapped to the problems of congestion control in VANET. The solutions provided by this algorithm consists of four components: 1) Transmission rate, 2) Transmission range, 3) Jitter, and 4) Delay.

The algorithm works as follows: The initial solution of MOTabu algorithm for congestion control is based on the current state, and it consists of current values of transmission rate, transmission range, delay, and jitter. Based on the initial solutions, the neighborhood set is generated. The generation of neighborhood set in MOTabu algorithm is based the values specified in the DSRC standard (i.e., transmission range – (10 – 1000 m) and transmission rate – (3 – 27 Mbps) [83-86]. Once the neighborhood set is generated, a feasible solution can be determined. Based on the feasible solution, the candidate list is generated and then searched for providing the best result.
MOTabu consists of three memory mechanisms: 1) Short-term memory, 2) Mid-term memory, and 3) Long-term memory and each mechanism has its functions [47]. Short-term memory mechanism is used to eliminate repeated solutions. It is based on tabu list (i.e., it contains a list of forbidden solutions).

The best solution identified from the candidate list is compared with tabu list values. If the solution is already present in the tabu list, the new solution is eliminated, or if the new solution is not available, it is declared as a new solution and it is inserted into the tabu list. The maximum size of the tabu list is fifty. The initial solution gets removed if the list is full. Mid-term mechanisms help to deepen the search based on specific areas of the solution space which is calculated based on the best solutions determined in the tabu list. Mid-term and long-term memory mechanisms are used for building optimal solutions and diversification of these solutions. Long-term memory mechanisms create various solutions by restarting the search process. The novel solutions should be determined when compared to previously generated solutions. This helps in avoiding entrapping in the local minima. Figure 9 shows the MOTabu memory mechanisms.

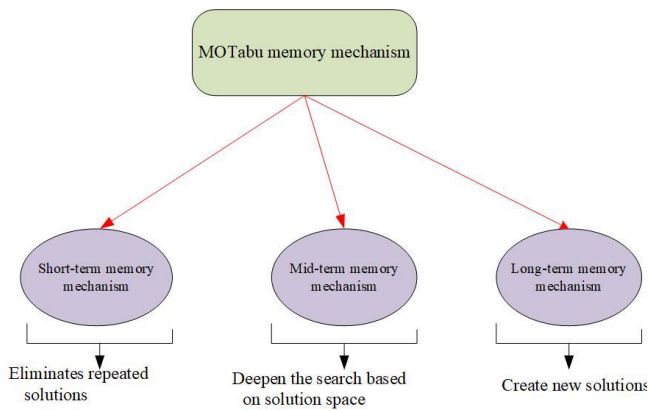

Figure 9: MOTabu memory mechanisms.

*ii) SPAV:* SPAV adjusts the transmission rate to reduce congestion in a connected vehicular environment [86]. After the congestion is detected on a network, the SPAV algorithm is executed on each vehicle to reduce the congestion. The SPAV algorithm reduces the congestion based on the position of each vehicle and communication channel used for transmitting the messages between the sender and the receiver. The authors proposed the Distributed Vehicle Density Estimation (DVDE) technique to acquire the position of the neighboring vehicles [87]. The positions of the neighboring vehicles are used as input to the SPAV algorithm to compute the transmission rate of each vehicle in a region and to calculate the maximum area in which vehicles are allowed to transmit beacons within a target threshold value to minimize the congestion. However, the proposed SPAV algorithm is suitable only for low vehicle densities and suffers from large end-to-end delays and high routing overheads at high vehicle densities such as downtown regions.

*4.1.3 MORS*

MORS is an overhead-free congestion control approach in which two primary metrics, i.e., Packet Reception Rate (PRR) and Distance over Communication Ratio (D/CR), which are measured at each node to reduce the overall delay due to reliability and fewer hops. It is operated in two different phases: 1) Fully Distributed Congestion Control (FD2C), 2) Unicast Multi-hop Data Dissemination (UM2D). Among these, FD2C guarantees on-hop message delivery, and UM2D performs node selection based on PRR and the D/CR ratio. The assumptions of these approaches are: 1) All vehicles are equipped with DSRC and use the Vehicular Deterministic Access (VDA) channel access scheme [88], 2) Same attenuation of signals for all vehicle directions, 3) Message size and frequency are the same for all nodes [89].

*i) FD2C:* The FD2C mechanism provides distant message delivery based on Communication Density (CD). It controls the load at each node by adjusting its transmission power. Based on CD at each node the transmission power can be reduced [90]. The reason behind reducing the transmission power at each node is based on two important factors 1) Higher transmission range affects the transmission of other nodes, 2) Higher transmission range leads to a higher detection range which in turn receives the traffic of all other nodes. But, reducing the transmission power results in high packet loss and collision rate at high traffic conditions.

*ii) UM2D:* When 802.11p was designed, the maximum distributed transmission range was 1000 m but in one-hop transmission, the maximum range we can reach is 300 m [91]. Although FD2C provides a higher transmission range, many nodes that fall within the range are selected as a relay. To address this problem, we need metrics to distinguish nodes from one another. Also, it is mentioned that a single metric cannot be used to solve this problem [92-95]. In the UM2D distance and link, quality is used to solve this problem and based on these two metrics multi-hop dissemination solutions are proposed. Link quality is directly proportional to PRR, (i.e., better link quality provides better reception rate). Distance is directly proportional to end-to-end delay (i.e., end-to-end delay is based on the number of hops), a smaller number of hops leads to a lower end-to-end delay. Hence, the distant node is chosen for communication [96]. The end-to-end delay of UM2D is low at all vehicle densities and traffic conditions. However, it requires high bandwidth to provide better transmission and reception rates.

**4.2 Open loop congestion control**

Open-loop congestion control mechanisms avoid congestion before it happens. In this paper, we discussed prioritizing and scheduling approach used in open-loop congestion control environment. The open loop congestion control strategies are based on three main strategies: 1) Priority based congestion control, 2) Topology based congestion control, and 3) Adaptiveness based congestion control. Table 3 presents a summary of open loop congestion strategies.

*4.2.1. Location and priority-based congestion control*

VANETs suffer from channel congestion in a highly dense situation, which leads to performance degradation. To improve performance, safety, and reliability, we study two strategies: 1) Dynamic Scheduling (DySch), 2) Static Scheduling (TaSch) [49]. These strategies assign priority to the messages based on message size, message content, and network usage. This approach is an open loop congestion control because the congestion control mechanisms are applied before congestion. DySch and TaSch consist of two different units in congestion control. The first unit is known as priority assignment unit. It is responsible for assigning priority to the safety messages generated based on static and dynamic factors. The second unit is known as message scheduling unit. It is responsible for rescheduling prioritized messages in control channel and service channel queues. Its function is different in both strategies. The priority assignment unit assigns priority to the message. The authors assumed that emergency and high priority messages sizes are smaller than normal messages [49]. As we mentioned earlier, the priority of the messages is determined based on static and dynamic factors. The static factor is identified based on the message content and the application type, and each beacon message can have priority ranging from 1 to 5. Beacon messages are commonly used to identify position, speed, and the direction of neighboring vehicles [46]. These messages play an important role in congestion control. The highest priority in static factor is priority 5 which is assigned to emergency messages. These messages must be sent without any delay. Intersection collision warning, pedestrian crossing, and vehicle approaching warning are some of the applications. The dynamic factor is identified based on parameters such as the speed of the vehicle, importance of the message, validity of the message, the distance between the sender and the receiver, and the direction of the sender and the receiver. It is calculated based on the GPS information and the routing table. In addition, the Enhanced Distributed Channel Access (EDCA) and Network Coding-aware Admission Control (NCaAC) algorithm is used to prioritize the messages. Figure 10 shows the approach of priority-based congestion control strategies.

**Table 3: Summary of open-loop congestion control methods based on different parameters.**

| Host centric technique | Approach | Topology | Adaptiveness | Simulator used | Channel used | Performance metrics |
|---|---|---|---|---|---|---|
| Pre-congestion | Prioritizing and scheduling message (DySch, TaSch) [49] | Decentralized | Adaptive | SUMO, MOVE and Network Simulator v2.35 | CCH, SCH | Event-driven |
| | Context Awareness Beacon Scheduling (CABS) [97] | Decentralized | Adaptive | Qualnet | CCH | Measurement driven |
| | EDCA [98] | - | - | NS2, OPNET modeler | - | - |
| | Improve route stability of AODV [99] | Decentralized | Adaptive | GloMoSim | - | Event-driven |
| | Adaptive congestion control [100] | Centralized | Adaptive | Event-driven simulator | CCH | Event-driven |
| | Cooperative and distributive [87] | Decentralized | Adaptive | UPPAL | CCH and SCH | Event-driven |
| | Cooperative Load transfer (CLB) [101] | Centralized | Adaptive | CDMA Subscriber Identity Module | Request and response channel | Measurement driven |
| | Stability-based AODV [102] | Decentralized | - | SUMO, MOVE and NS2 | Wireless channels | Event and measurement driven |
| Post-congestion | Dynamic bandwidth [64] | Centralized | - | - | CCH and SCH | Event-driven |
| | AODV location-based routing scheme [77] | Decentralized | Adaptive | Network Simulator (NS2.34) | Wireless channels | Event-driven |
| | LBV-AODV routing [115] | Decentralized | - | Network Simulator (NS2.34) | Wireless channels | Event-driven |
| | Stability-based approach [91] | Centralized | - | OPNET-16.0 | CCH and SCH | Event-driven |
| | DCC [104] | Decentralized | - | SUMO and NS2.34 | - | Event-driven |
| | Limeric [104] | Decentralized | Adaptive | NS2 | - | Event-driven |

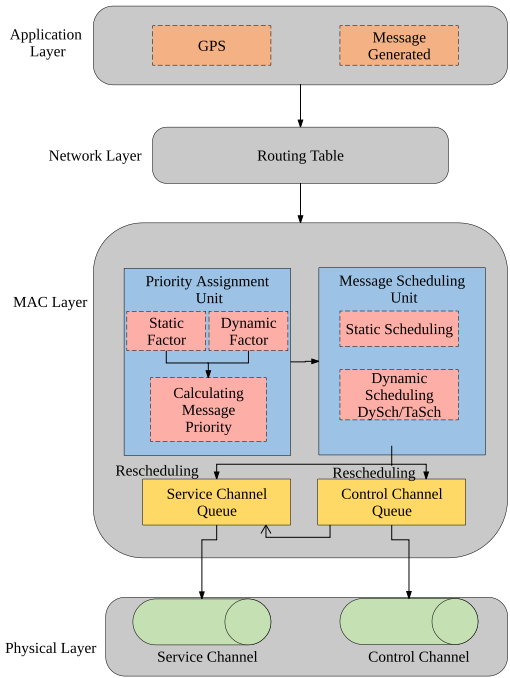

**Figure 10: Congestion control approach in prioritizing and scheduling messages.**

*i) DySch:* In DySch, the vehicle that travels at high speed is high priority when compared to other vehicles due to the probability of disconnection being high. The importance of the messages is calculated based on the ratio of the total communication area and the overlapped area [105]. When the overlapped area is very high, the importance of the message is low due to the high probability of a message being received again from nearby vehicles. Thus, messages with lower use metric are lower priority messages. The validity of messages is calculated based on the ratio of the lifetime of a message to its transmission time. If the lifetime of the message is very low, it indicates that the message must be delivered without any further delay.

Thus, high priority is assigned to the messages with a lower lifetime [91]. The greater the distance between the sender and the receiver, the higher is the disconnection probability. Hence, vehicles with larger distances have higher priorities. The direction of the sender and the receiver represents the directions the sender and the receiver are traveling. If the sender and receiver are traveling in the same direction, the probability of connection increases resulting in a low priority connection. Static and dynamic factors are used in calculating priorities. The calculated priorities are embedded in message headers.

*ii) TaSch:* In Tasch, the message scheduling unit is used for rescheduling prioritized messages into control channels and service channels. It is a challenging task in VANET due to its dynamic topology, high mobility, high vehicle speed, and so on. Similar to the priority unit, the message scheduling unit is also divided into two main categories: 1) Static scheduling, 2) Dynamic scheduling to transfer queues present in the control channel and service channel before rescheduling takes place.

In static scheduling, based on the priorities of messages, they are delivered to the control channel or service channel [49, 106]. Messages with high priorities are transferred using the control channel queue, and message with low priorities are transferred using the service channel. If the control channel queue is full, messages with high priorities are transferred to service channel queues since these messages need to be delivered without any delay. Dynamic scheduling is accomplished using two different methods: 1) Predefined message priority, 2) Meta-heuristic techniques for rescheduling queues. In the predefined message, priority messages are prioritized by the priority assignment unit. When new packets arrive, packets inside each queue are rescheduled. Then, these packets are de-queued to service channel queues or control channel queues to be transferred over the control channel or the service channel. Figure 11 and figure 12 show the control channel and service channel queues respectively.

*iii) EDCA:* EDCA prioritizes messages transmitted over a control channel for safety messages and six service channels for non-safety messages. The sender's relative speed is related to the speed of the other vehicles in that region, which is calculated based on the distance the vehicle covers over a given time interval (*t*) and then normalized by using the communication range of a vehicle [98].

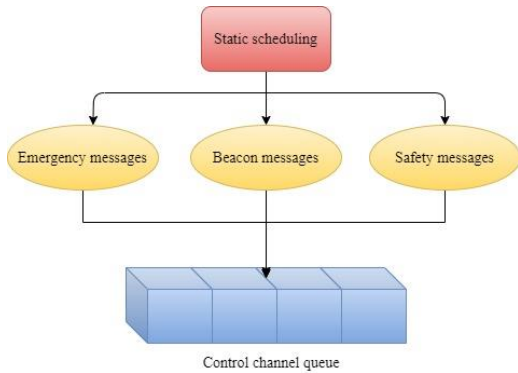

**Figure 11: Control channel queue.**

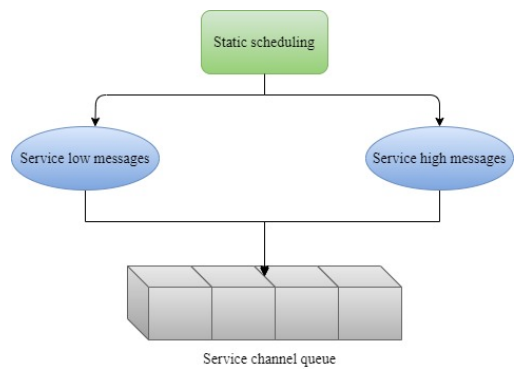

**Figure 12: Service channel queue.**

**Table 4: Evaluation results/findings**

| Approach | Delay | Overhead | Packet loss | Energy | Mobility |
| --- | --- | --- | --- | --- | --- |
| Prioritizing and scheduling message [49] | High | Medium | Low | Medium | High |
| CABS [97] | Medium | High | Medium | Low | Medium |
| EDCA [98] | Medium | High | Low | High | Medium |
| Improve route stability of AODV [99] | Low | Low | Medium | High | Medium |
| Adaptive approach [100] | High | Medium | Medium | Low | High |
| CLB [101] | Medium | High | High | Medium | High |
| LBV-AODV routing [115] | Low | High | High | Medium | Low |
| DCC [104] | High | High | Low | High | Medium |
| Cross layer approach [61] | Medium | Medium | High | Low | Medium |
| DD-FPAV [79] | Low | Low | High | High | Low |
| UOTabu [47] | Low | Medium | Low | Medium | High |
| MOTabu [47] | High | Medium | Medium | High | Low |
| Multistate active DCC [66] | Medium | Medium | Low | High | High |
| Limeric [104] | High | Medium | Low | Low | High |
| CARTIM [65] | Medium | Low | Low | High | Medium |
| PULSAR [71] | High | Medium | Low | High | Medium |
| COMPASS [70] | Medium | Medium | High | Medium | High |
| DSRC based congestion control [72] | High | Medium | Medium | Low | Medium |

The EDCA algorithm provides high channel access to the high priority messages such as safety messages transmitted over a network, which further helps in controlling the message transmission rate to reduce the congestion in a connected vehicular environment [107]. Moreover, EDCA supports Carrier Sense Multiple Access with Collision Avoidance (CSMA/CA), the EDCA algorithm senses the medium before transmitting the messages to avoid the collision of packets. The limitations of EDCA include high overhead and energy arising from additional information embedded in each packet to detect congestion in the network.

*iv) NCaAC:* In NCaAC [108], network coding techniques are applied to reduce congestion among connected vehicles. The RSU classifies messages based on high and low priorities and then, the high and low priority messages transferred to the control channel and service channel queues, respectively. If the number of vehicles increases in the region or at high traffic conditions, the RSU performs load balancing with nearby RSUs to minimize the occurrence of network congestions. The benefits of network coding for congestion in VANET are efficient bandwidth utilization and low packet loss at all vehicle densities. However, NCaAC suffers from high energy consumption and high overhead while performing load balancing to redistribute the loads to nearby RSUs.

*4.2.2. Topology-based congestion control*

Topology-based congestion control is based on centralized and decentralized approaches. Centralized Congestion Control approaches assume a central controller such as RSUs to control the signal parameters and path information to guide the vehicles. The RSUs and OBUs direct all DSRC connected vehicles to provide on-demand information about the ongoing network traffic such as speed, position, acceleration, braking status, etc. of the neighboring vehicles [109-112]. Centralized approaches are easier to implement because they incur less overhead in routing connectivity. Common examples of centralized approaches for congestion control include robust congestion control scheme [67, 111], dynamic sharing of bandwidth approach [113], dynamic congestion control approach [100], and CLB approach [101].

The decentralized or distributed congestion control approach is the default for VANETs. In this approach, there is a set of local controllers distributed in the entire network and each local controller can extract only limited information of beacon message such as speed, concentration, and travel time

MOTabu [47, 112], Improvised AODV [114], LBV_AODV [115], and CABS [97] are examples of decentralized approaches for congestion control that have been proposed for VANETs.

*i) Robust congestion control:* Robust congestion control scheme is one of the topology-based centralized congestion control approaches that minimizes congestion in the VANETs environment [67, 111]. The congestion detection center detects congestion in the network based on the priority level and the number of hops the messages have traveled with using metrics such as average message waiting time, collision rate, and message reception rate. The congestion control center minimizes the congestion by adjusting transmission power, as well as the message generation and transmission rate. The advantages of the robust congestion control scheme include low overhead and end-to-end delay. However, this approach suffers from a high packet loss when the number of vehicles increases in the region.

*ii) Dynamic sharing of bandwidth:* In this approach, the congestion detection mechanism monitors the channel status, priority of a node, and message queue length to detect congestion among connected vehicles [113]. The RSU computes the priority of messages generated by each vehicle based on message content, size, and transmission time and allocates bandwidth dynamically depending on the message priority. The congestion control mechanisms are applied before congestion occurs in the network, and if any of the service channels are overloaded, the RSU transfers the messages from the service channel queue to the control channel queue to minimize the congestion and message transmission delay, which lowers end-to-end delay and packet loss. However, some of the shortcomings of dynamic bandwidth sharing include high overhead and energy consumption which arise from transferring the messages from the service channel queue to the control channel queue.

*iii) Dynamic congestion control:* Dynamic congestion control is the most commonly used topology-based centralized congestion control approach in VANETs. The centralized controller (i.e., RSU) calculates the possible message transmission rate based on the number of vehicles and then transmits the information to all the vehicles in the region. Upon receiving the message from the RSU, each vehicle modifies the configuration of the message transmission rate to the value specified by the RSU, which helps in minimizing congestion in the network even at high vehicle density regions. Moreover, dynamic congestion control provides high availability and channel capacity for high priority messages and maximum channel utilization for low priority messages [100]. The benefits of this approach are low overhead and low packet loss. But the end-to-end delay of the dynamic congestion control approach increases slightly when the number of vehicles increases because the RSU needs to monitor each vehicle to generate the message transmission rate.

*iv) CLB:* In CLB, the authors performed a load balancing technique to reduce congestion in the network. The proposed algorithm schedules the messages to be transmitted using the Earliest Deadline First (EDF) and Slack Time Inverse (SIN) scheduling algorithms, and RSUs broadcast the messages to all vehicles in the region. RSUs are deployed only in the critical regions of VANETs [101]. Moreover, RSUs have limited bandwidth and coverage area, which results in overloaded RSUs at high vehicle densities. If any of the RSUs are overloaded with more messages, the congestion control center in CLB applies a load balancing technique to redistribute the loads to nearby RSUs to minimize the congestion and packet loss. This results in higher end-to-end delay and overhead at high vehicle densities compared to robust congestion control [67, 111] and dynamic bandwidth sharing [113] techniques.

*v) MOTabu:* MOTabu is one of the most commonly used dynamic and topology-based decentralized congestion control techniques in VANETs. The MOTabu algorithm for congestion control is based on the current state of the vehicle and uses the following performance metrics: transmission rate, transmission range, delay, and jitter [47, 112]. Section 4.1.2 describes the operation of the MOTabu congestion control approach.

*vi) Improvised AODV:* In the improvised AODV approach, the congestion control algorithm monitors the channel status, and service and control channel queue capacity to detect congestion in the region. If congestion occurs, the congestion control algorithm selects a leader vehicle with high available resources and buffer capacity [114]. The leader vehicle transmits a new message transmission rate to all other vehicles in the region to reduce the load and bandwidth utilization in the network. The advantages of the improvised AODV approach include low overhead and low collision ratio. However, the end-to-end delay and packet loss increase whenever congestion occurs due to limited available resources and bandwidth.

*vii) LBV_AODV:* This approach applies a load balancing technique to the vehicles to reduce congestion. The congestion detection in the LBV_AODV algorithm calculates and monitors the queue length of all vehicles to detect congestion in the network [115]. The load balancing technique is applied to redistribute the load from overloaded vehicles to the least loaded vehicles to minimize the congestion. Moreover, if there is any broken link, an alternate path is chosen to transmit the messages to the destination vehicle. The LBV_AODV approach results in high overhead, end-to-end delay, packet loss at high vehicle density regions because a large number of messages need to be delivered within a specific time interval. The benefits of LBV_AODV congestion control are high throughput and low end-to-end delay at low vehicle densities.

*viii) CABS:* CABS is a decentralized distributed beacon scheduling technique which minimizes congestion in VANETs. The authors used the Time Division Multiple Access (TDMA) technique to schedule the beacon messages with a predictable delay and high reliability to control the channel load and capacity [97]. In CABS, every vehicle gets different timeslots to transmit the messages to other vehicles based on the Virtual Time Frame Table (VTF). VTF contains vehicle information, time slot, and transmission rate, shared among all the vehicles in the region. If a vehicle misses the timeslot, it has to wait for the next time slot to transmit the messages, which reduces the collision of packets to a greater extent compared to MOTabu [47], Improvised AODV [114], and LBV_AODV [115] techniques. Other vehicles can use the reserved timeslot if they encounter emergency situations (such as a road accident, brake failure) and the VTF table is modified accordingly to notifies the changes to the other vehicles. However, this approach suffers from high end-to-end delay when the number of vehicles increases in the system because the vehicles have to wait for their timeslots to transmit messages to other vehicles.

*4.2.3. Event-driven and adaptive-based congestion control*

The event-driven and adaptive congestion control approach dynamically changes the control decisions in the topology as well as the network traffic when congestion occurs in the channel. The control decisions are dynamically changed as a function of the current network state [116]. The approach guarantees robustness, convergence, and stability of closed-loop systems. The non-adaptive approach does not change the control decisions in the topology based on network traffic, load, routing, and so on. The approach is fixed and static and does not consider the current state of the network. The non-congested node does not know the status of the other nodes. Speed Based Adaptive Power Control (SBAPC) [117], Beacon inter-reception time Ensured Adaptive Transmission (BEAT) [118], adaptive strategy congestion control [119], and Adaptive transmit power Cooperative Congestion Control (AC3) [120] are most commonly used adaptive-based congestion control schemes in VANETs.

*i) SBAPC:* SBAPC dynamically changes control in the topology when congestion is detected in the vehicular network [117]. In SBAPC, each vehicle dynamically adjusts the transmission rate and transmission power of BSM messages based on vehicle speed, position, and channel congestion. The main objective of SBAPC is that Time To Collision (TTC) with neighboring vehicles decreases when the speed of the vehicle increases. Like SPAV (Section 4.1.2), the position of the neighboring vehicles is used as input in the SBAPC algorithm to compute the transmission rate of each vehicle in a region and to calculate the maximum area in which vehicles are allowed to transmit beacons within a target threshold value to minimize the congestion. However, SBAPC suffers from high end-to-end delays and communication overheads.

*ii) BEAT:* In this approach, the authors proposed a congestion control based on the beacon transmission and reception rate [118]. The congestion detection includes parameters such as channel usage level, channel occupancy time, and the number of messages in a queue. To minimize the congestion in the network, the BEAT framework varies the beacon message generation based on vehicular safety, the density of vehicles, and bacon reception. The advantages of the BEAT framework include high availability and channel capacity for high priority messages and maximum channel utilization for low priority messages. However, the end-to-end delay and packet loss increases whenever congestion occurs due to limited available resources and bandwidth.

*iii) Adaptive strategy congestion control:* In [119], the proposed adaptive congestion control reduces the congestion based on the surrounding road traffic conditions. The RSUs monitors and calculates the density of the vehicles in that region. Hence the channel loads are distributed to nearby RSUs when the number of vehicles increases to balance the load on the network. The proposed adaptive congestion control [119] minimizes the congestion even for high vehicle density regions. However, limited bandwidth results in high overhead and energy consumption while balancing the load on the network.

*iv) AC3:* In AC3, the authors performed adaptive congestion control strategies to reduce channel congestion and maximize individual payoffs based on the principles of game theory. The AC3 algorithm allows the vehicle to select the transmission rate and transmission power autonomically based on the density of vehicles in a region and the channel usage [120]. The AC3 algorithm requires vehicles with the highest contributions such as vehicles with high transmission rates and transmission power as input to reduce the congestion. The mathematical model is used to calculate the transmission rate, transmission power, speed, and position of the vehicles in a region. From the simulation results of AC3, we can infer that AC3 is one of the robust and efficient adaptive congestion control strategies to reduce channel congestion in VANETs.

Table 4 presents various recently proposed congestion detection and control protocol based on various characteristics such as delay, overhead, packet loss, energy, and mobility. There is always a tradeoff when choosing an appropriate congestion control and congestion detection approach. Based on our literature review, table 4 provides pertinent information to help a designer choose the most suitable protocol. We looked at the details in the paper including mathematical expressions, and data/results to come up with the classification of the delay, overhead, packet loss, energy, and mobility into low, medium, and high.

In conclusion, section IV discussed both closed-loop and open-loop congestion control techniques for connected vehicles, out of which, the majority of the closed congestion control techniques have limitations, such as, higher end-to-end delay, packet loss, and collision ratio compared to open-loop congestion control techniques which apply congestion control mechanisms after the congestion is detected [sentence did not make sense – check if this is what you mean else rewrite it]. In contrast, recently proposed open-loop congestion control are used widely to avoid congestion before it occurs as VANETs are highly dynamic, and failure to deliver the messages interrupts communication between vehicles resulting in catastrophic consequences. The open-loop congestion control techniques monitor the network continuously, and if there is a chance of congestion, the proposed congestion control algorithms act immediately to minimize the possibility of congestion occurring in the region. However, some of the open-loop congestion control techniques suffer from high overhead and delay due to the large number of instructions exchanged between the vehicles to reduce load and bandwidth utilization of the network.

We have also discussed the challenges of the congestion control techniques in Section V, which gives further insight to the readers and researchers about the issues that still exist in current congestion control techniques. Moreover, to overcome the drawbacks of existing congestion control techniques and to perform congestion detection and control efficiently, we have recommended existing state-of-the-art techniques in future directions.

## V. CHALLENGES AND FUTURE DIRECTIONS

Even though remarkable progress has been made in the field of connected vehicles and specially in the area of congestion control they are still insufficient to meet the dynamic requirements of VANETs in realistic solutions. Many techniques which work fine in other traditional network settings cannot be directly deployed to the connected vehicle environment because of the many characteristics that are unique to this network. In previous sections, we have discussed several techniques used for congestion detection and control among connected vehicles. In this section we discuss some of the concerns and possible future research directions.

### 5.1. Computational Intelligence

Building computational intelligence to manage congestion control and identification is an interesting approach in the field of connected vehicles. Machine Learning (ML) is being heavily used in various industries due to its capability of learning from data provided by various algorithms and finding solutions to various problems. It has been used successfully for congestion control in wireless network environments [121, 122]. Machine learning base congestion control and identification approaches are based on either supervised and unsupervised learning. But in the field connected vechiles, which has a dynamic topology, traditional approaches may not be sufficient. Possible solutions can be provided by reinforment and deep learning.

*i) Reinforcement learning:* It is a subfield within ML, which has a huge potential to solve the congestion identification and control problem associated with the dynamic environment and develop satisfactory policies to meet diverse QoS requirements of vehicular networks while adapting to the varying wireless environment. For example, in resource allocation problems, the optimal policies are first learned and then the vehicle's agents accordingly take actions to adjust power and allocate channels adaptively to the changing environments characterized by, e.g, link conditions, locally perceived interference, and vehicle kinetics while traditional static mathematical models are not good at capturing and tracking such dynamic changes [123].

*ii) Deep learning and High Performance Computing (HPC):* Deep learning aims to learn data representations which can be built in supervised, unsupervised, and reinforcement learning and has made significant advances in various machine learning tasks. Unlike traditional machine learning techniques that require a lot of efforts on feature design, deep learning provides better performance by learning the features directly from raw data [124]. Deep learning is a deeper version of neural networks, which consists multiple layers of neurons. Each neuron in the network performs a non-linear transform on a weighted sum of a subset of neurons in its preceding layer. However, numerous several challenges (such as much more training data is needed) exist when training deeper networks. Moreover, with high- performance computing facilities, such as

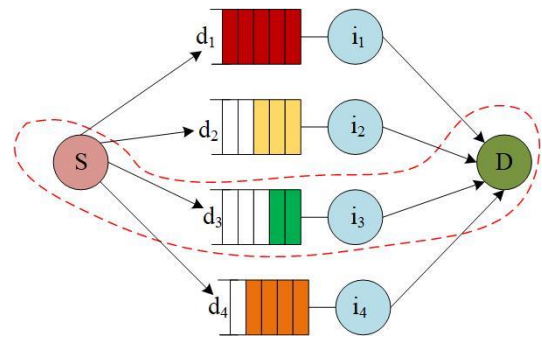

**Figure 13: Markovian approach.**

Graphics Processing Unit (GPU), deep networks can be efficiently trained with massive amounts of data through advanced training techniques such as batch norm and residual networks [124].

Although has advances have been made in the field of artificial intelligence but, solving the congestion control and identification problem in a vehicular network remains a significant challenge. Standalone ML techniques may not be enough to provide the solution. We have to combine ML techniques so that computational complexity can be more efficiently managed.

### 5.2. Markovian Routing

The Markovian routing approach combines Markovian Traffic Equilibrium (MTE) with the Network Utility Maximization (NUM) model to form a Markovian Network Utility Maximization (MNUM) [125]. It establishes the uniqueness of equilibrium which allows providing a design for congestion control with multipath routing. Here, the source node generates link estimated delays which yield an end-to-end delay for each network node. In this approach, end-to-end delays are calculated based on the MNUM and packets are routed based on the node which has the least delay. It is also possible to adjust the flow of information in each link based on the delay information. Figure 13 shows the Markovian routing approach. Here S and D are the source and destination nodes, respectively, $i_1$, $i_2$, $i_3$, and $i_4$ are the intermediate nodes, $d_1$, $d_2$, $d_3$, and $d_4$ are the end-to-end delays of intermediate nodes $i_1$, $i_2$, $i_3$, and $i_4$, respectively. The link with the lowest delay (i.e., $i_3$) is chosen to route the packets from source to destination. The Markovian routing technique can be applied to RSUs to reduce end-to-end delay arising from collecting the information about vehicles resulting in low end-to-end delay at all vehicle densities, including regions with high density of vehicles.

### 5.3. Location awareness

Congestion of networks can be controlled in connected vehicles using location awareness. The use of a location awareness application makes it possible to determine a user's current location and future location. Based on this information we can classify the users and establish a connection among them [126]. If the number of users for a particular location increases, congestion will occur. To avoid this scenario a vehicular cloud can be established which can provide viable services to them. Location awareness is different from most existing machine learning applications that assume easy availability of data. However, in vehicular networks, the data is generated and stored across different units in the network, e.g., vehicles, RSUs, and remote clouds. Consequently, distributed learning algorithms are desired so that they can act on partially observed data and meanwhile have the ability to exploit information obtained from other entities in the network. Such scenarios can be technically modeled as a multi-agent system, where cooperation and coordination among participating agents play important roles in achieving optimal performance at the system level by sharing the necessary information among each other. Thus, each individual vehicle's agent is more aware about the environment and jointly optimizes its performance along with other agents in the network. With machine learning, vehicle agents are able to learn what they need to share based on what they have perceived and what they need to do, with minimal network signaling overhead.

## VI. CONCLUSION

In VANET, congestion detection and control are the two crucial factors which need to be considered because they have a direct impact on the performance of the network. In this paper, we have reviewed recently proposed techniques for congestion detection and control in VANETs. After reviewing various aspects of existing approaches such as routing protocol, technique used, type of congestion control, we have identified drawbacks of these existing approaches. We have summarized the features and characteristics of different congestion control approaches used for VANETs. We have categorized these approaches based on several criteria. We have also suggested existing solutions such as Markovian routing, statistical approach, machine learning, and location awareness to overcome the drawbacks of existing techniques as well as to provide more efficient services to users.

Thus, to enhance the safety of users, new research programs, protocols, architectures need to be determined and developed. Hence, we need substantial assistance from the government and automotive industry and a concrete endeavor between the academic community and industry. VANET is the next technological paradigm that guarantees safe road travel because of its tremendous potential to reduce road accidents and increase the safety of passengers.

## VII. ACKNOWLEDGMENTS

We thank the anonymous reviewers for their valuable comments which helped us improve the content, organization, and presentation of this work.

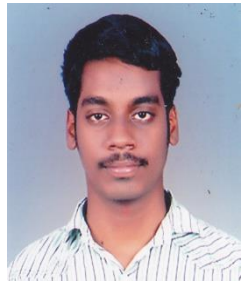

**Anirudh Paranjothi** received the B.E. degree in computer science from Anna University, Chennai, India in 2014 and his M.Sc in computer science from Texas A&M University, Kingsville in 2016. He is currently a Ph.D. student at the University of Oklahoma. His research interests are Vehicular Ad-hoc Networks (VANET), social networks, Computer networks, Mobile cloud computing and Software engineering.

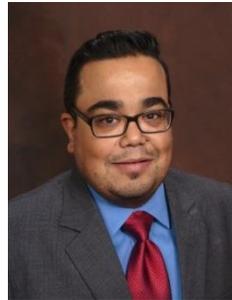

**Mohammad S. Khan** ((SM' 19) is currently an Assistant Professor of Computing at East Tennessee State University and the director of Network Science and Analysis Lab (NSAL). He received his M.Sc. and Ph.D. in Computer Science and Computer Engineering from the University of Louisville, Kentucky, USA, in 2011 and 2013, respectively. His primary area of research is in ad-hoc networks, wireless sensor networks, network tomography, connected vehicles and intelligent transportation systems. He currently serves as an associate editor of *IEEE Access, IET WSS, IET ITS and Springer Telecommunications System.* He has been on technical program committees of various international conferences and technical reviewer of various international journals in his field. He is a senior member of IEEE.

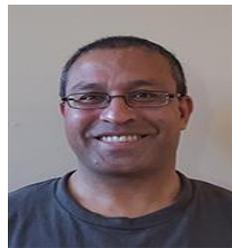

**Sherali Zeadally** earned his bachelor's degree in computer science from the University of Cambridge, England. He also received a doctoral degree in computer science from the University of Buckingham, England, followed by postdoctoral research at the University of Southern California, Los Angeles, CA. He is currently an Associate Professor in the College of Communication and Information, University of Kentucky. His research interests include Cybersecurity, privacy, Internet of Things, computer networks, and energy-efficient networking. He is a Fellow of the British Computer Society and the Institution of Engineering Technology, England.